\documentclass[manuscript]{aastex}

\begin{document}
\pagenumbering{arabic}

\title{THE DWARF SATELLITES OF M31 AND THE GALAXY}

\author{Sidney van den Bergh}
\affil{Dominion Astrophysical Observatory, Herzberg Institute of Astrophysics, National Research Council of Canada, 5071 West Saanich Road, Victoria, BC, V9E 2E7, Canada}
\email{sidney.vandenbergh@nrc.gc.ca}

\begin{abstract}

The satellite systems of M31 and the Galaxy are compared.  It is noted that all five of the suspected stripped dSph cores of M31 companions are located within a projected distance of 40 kpc of from the nucleus of this galaxy, whereas the normal dSph companions to this object have distances $>$ 40 kpc from the center of M31. All companions within 200 kpc $<$ D(M31) $<$ 600 kpc are late-type objects. In one respect The companions to the Galaxy appear to exhibit different systematics with the irregular LMC and SMC being located at small $R_{gc}$. It is speculated that this difference might be accounted for by assuming that the Magellanic Clouds are interlopers that were originally formed in the outer reaches of the Local Group.  The radial distribution of the total sample of 40 companions of M31 and the Galaxy, which is shown in Figure 1,  may hint at the possibility that these objects contain distinct populations of core (R $<$ 25 kpc) and halo (R $>$ 25 kpc) satellites.)

\end{abstract}

\keywords{galaxies: dwarf - galaxies: individual (M31, Galaxy)}

\section{INTRODUCTION}

In the present investigation the data on the companions to M31 and the Galaxy are extended by including a number of recently discovered satellites. Furthermore, following Koch \& Grebel (2006), some compact objects that are widely believed to be the stripped cores of now defunct dwarf spheroidal galaxies, have been added to the list of satellites to M31 and the Galaxy. This enlarged database is then used to investigate some of the systematics of the
M31 and Milky Way satellite systems. In particular we seek to answer three questions: (1) How does the morphological type of a satellite depend on its distance from the center of its parent galaxy? (2) Do inner and outer dwarf satellites belong to separate core and halo populations, and (3) were the Magellanic Clouds formed as satellites of the Galaxy, or might they have been  captured from the outer reaches of the Local Group? 

\section{ DATA ON LOCAL GROUP SATELLITES}

\subsection{\it The companions of M31.}

A listing of the known companions to the Andromeda galaxy is given in Table 1. Also included in this table are five objects that lie above the relation of Mackey and van den Bergh (2005), and which might therefore be regarded as candidate stripped cores of dwarf galaxies, rather than as extended globular clusters. The distances to, and structural parameters of, the majority
of the M31 companions were taken from the recent compilation by McConnachie \& Irwin (2006). Following Ferguson, Gallagher \& Wyse (2000) we assume that And IV is, in fact, a background galaxy that is viewed through the disk of the Andromeda galaxy. Furthermore the reality of Andromeda VIII (Morrison et al.
2003) does not yet appear to have been firmly established (Merrett
et al. 2006). This object has therefore been excluded from the present compilation.  Also Andromeda NE (Zucker et al. 2004) has been omitted from the
sample because the true nature of this object is still not firmly established. However, following this same reference , Andromeda IX has been included in Table 1. Andromeda X (Zucker et al. 2006) was also accepted as a dwarf spheroidal companion to M31. Also included in Table 1 are Mayall II = G1, which is suspected of being the stripped core of a dwarf galaxy (Meylan et al. 2001) and the most luminous M31 globular cluster 037-B327, which is also believed to be the stripped nucleus of a dwarf spheroidal (Ma et al. 2006). Also
included in Table 1 are the three extended objects discovered recently by Huxor et al. (2005) which lie above and to the left of the relation of Mackey and van den Bergh (2005), that appears to separate true globular clusters from the stripped cores of dwarf spheroidals. All five of the objects discussed above appear projected close to the nucleus of the Andromeda galaxy and were therefore assumed to be situated at the same distance from the Sun as M31 itself.  Both proper motions and radial velocities will be required to establish which of the objects listed in Table 1 are true satellites of M31, and which ones should more properly be regarded as free-floating members of the Andromeda subgroup of the
Local Group (van den Bergh 2000, p.285). Inspection of the data in Table 1 shows a clear dependence of morphological type on distance from the center of M31; an effect first noticed by Einasto et al. (1974). All 15 Andromeda satellites that are situated at D(M31) $<$ 200 kpc are seen to be of early type. On the other hand four out of seven of the galaxies with 200 kpc $<$ D(M31) $<$ 600 kpc
are of late type. [M33 (Sc), IC 10 (Ir), IC 1613 (Ir) and Pegasus = DDO 216 (Ir)]. 

\subsection{\it Companions to the Galaxy.}

A listing of possible physical companions to the Milky Way
system (van den Bergh 2000, Grebel, Gallagher \& Harbeck 2003) is given in Table 2. The object Willman 1 (Willman et al. 2006) has been excluded from the table because, with $M_{v} = -3.0$ and $R_{h} = 23$ pc, it falls well below the empirical relation (Mackey \& van den Bergh 2005) that appears to separate globular clusters from dwarf spheroidal galaxies. On the other hand the newly discovered Ursa Major system (Willman et al. 2005), which has $M_{v} = -6.75$ and $R_{h} = 250$ pc, lies far above this relation in the domain occupied by normal dwarf spheroidal galaxies. The Ursa Major system has therefore been included among the Galactic satellite galaxies listed in Table 2. On the other hand the Canis Major system was excluded because the possibility that it is a peculiar deformation or asymmetry of the outer Galactic disk cannot yet
be definitely ruled out (Bellazzini et al. 2006). Recently Carraro et al. (2006) have also suggested that NGC 6791 might be the nucleus of a tidally disrupted metal-rich galaxy. A more plausible suggestion (van den Bergh 2000, pp. 54-55)
would seem to be that NGC 6791 is a metal-rich open cluster that was ejected from the Galactic nuclear bulge by interactions with  the bar near center of the Galaxy. The fact that Carraro et al. find no significant abundance spread among the members of NGC 6791 also militates  against the suggestion that this object is the remnant core of a once more massive Galactic satellite. 

\section{DISCUSSION}

Inspection of the data in Table 1 and Table 2 shows that the Galactic satellite system differs from that of M31 in three important ways: (1) All inner satellites of M31 are early-type objects. On the other hand the LMC and the SMC are presently situated at small Galactocentric distances. This perhaps encourages the speculation (Byrd et al. 1994) that the Magellanic Clouds might be interlopers that were initially formed in a more remote region of the Local Group. [A recent paper on the orbit of the LMC (Pedreros et al.
2006) {\it assumes} that the LMC is gravitationally bound to, and in an elliptical orbit around, the Galaxy.]  (2) All of the suspected stripped cores in M31 occur at small (R $<$ 40 kpc) distances from the nucleus of M31. However, among companions to the Galaxy the putative stripped core NGC 2419 is located quite far ($R_{gc}$ = 92 kpc) from the Galactic center. This suggests that this object may have had a different evolutionary history from those of NGC 5139  = $\omega$ Centauri $(R_{gc}$ = 6 kpc) and NGC 6715 = M54 $(R_{gc}$ = 19 kpc). (3) McConnachie \& Irwin (2006) have drawn attention to the fact that the dwarf spheroidals associated with the Galaxy have half-light radii that are two or three times larger than those of the dSph galaxies surrounding M31. The recently discovered Galactic satellite in Ursa Major (Willman et al. 2005) strengthens and confirms this result. It is not yet clear if the observed systematic differences between the dSph satellites of M31 and the Galaxy are due to stronger tidal striping of Galactic companions, or if the presently available data sample might have been more strongly biased against the discovery of Galactic companions of low surface brightness. It should of course be emphasized that it is very likely that many additional very low luminosity satellites of both M31 and the Galaxy remain to be discovered.  With the discovery of the UMa system one finds that the satellites of M31 and of the Galaxy, that are located within 150 kpc of their parents, now have a spread in surface brightness in excess of 5 mag arcmin$^{-2}$ . On the other hand the more distant satellites Leo I, Leo II, And II, And VI and And VII appear to have a much smaller luminosity dispersion and all have a surface brightness higher than 25 mag arcmin$^{-2}$. This difference might be due (McConnachie \& Irwin 2006) to a radial surface density gradient, or perhaps more plausibly, to observational selection effects that have biased the sample of dwarf spheroidal satellites against the discovery of distant low surface brightness objects.  

In their comparison of (fossil) satellites with detailed numerical simulations of galaxy survival Gnedin \& Kratsov (2006) note a discrepancy between theory and observation, in the sense that the observed radial distribution of fossils shows an excess of satellites at small radial distances. The present data increase the size of this discrepancy because of the inclusion of the putative stripped cores of dwarf spheroidals which (with the sole exception of NGC 2419) are all located at quite small radial distances from the nuclei of M31 and the Galaxy.

A plot of the cumulative radial distribution of all of the satellites of M31 and the Galaxy is shown in Figure 1. This figure appears to show an
abrupt break at R $\sim$25 kpc. The existence of this sharp discontinity 
suggests that the six innermost satellites (B327 - 3 kpc, M32 - 6 kpc,
NGC 5139 = $\omega$ Centauri -6 kpc, Hux C1 -13 kpc, Hux 3 -14 kpc,
and Sgr -19 kpc) might, in some way that is presently not understood,
differ from the other satellites of the Galaxy and M31. The observed
excess of satellites at small galactocentric distances is surprising
because one  would actually have expected disruptive tidal forces to have
produced a deficiency of satellites with pericentric radii $<$ 30 kpc
(Gauthier, Dubinski \& Widrow 2006). It would be interesting to
know if the apparent existence of an excess population of dwarfs 
at small radial distances is related to a result of recent N-body
simulations (Lu et al. 2006) which appear to show that the assembly 
of cold dark matter halos occurs in two phases: (1) a fast-accretion stage
with a rapidly deepening potential well, and (2) a  slow-accretion
stage characterized by a gentle addition of mass to the outer halo
with little change to the inner potential well.

A Kolmogorov-Smirnov test shows no statistically significant
differences between the distributions of the galactocentric
distances of the companions of M31 and of the Galaxy. This
conclusion is consistent with that of McConnachie \& Irwin (2006)
which was, however, based on a smaller data sample. 

Within the, admittedly limited, accuracy of published
metallicity values there is no obvious systematic difference
between the Mv versus [Fe/H] relationships for the late-type
satellites of M31 and of the Galaxy.

In summary it appears that the M31 and Milky Way satellite
systems are broadly similar, except for the presence of the
LMC and the SMC, which might be interlopers that originated
in distant reaches of the Local Group. To check on this possibility
by detailed orbit computations one would have to have a much
improved knowledge of the three-dimensional shape and radial profile of
the gravitational potential of the Milky Way dark halo.

It is a pleasure to thank Ken Freeman, Eva Grebel, Nitya 
Kallivayalil and Mario Pedreros for helpful exchanges of
of correspondence. I am also indebted to an unusually helpful
anonymous referee.

\begin{deluxetable}{lllcrrr}
\tablewidth{0pt}   
\tablecaption{Dwarf companions to M31}
\tablehead{\colhead{Name}  & \colhead{Type}  &\colhead{RA~~~~~~~~~~Dec} & \colhead{R}  &\colhead{M$_{v}$} &\colhead{D(M31)} &\colhead{[Fe/H]}\\ &&\colhead{(J2000)} &\colhead{(kpc)} &&\colhead{(kpc)} &\colhead{(dex)} }

\startdata

B327  &   GC  &  00$^{h}$ 41$^{m}$ 35$^{s}$ +41$^{o}$ 14'55" &  785  & -11.7  &  3   &  ...
\\

M32   &   E2,N &  00  42  42  +40  51 55 &  785  & -16.5  &  6 &  -1.1
\\

Hux C1   &GC  &   00  38  20  +41  47 15 &  785 &  -7.1  &  13 &  ...  \\

Hux C3 &  GC  &  00  38  05  +40  44 39  &  785 &  -7.1 &   14 &  ...
\\

G1   &    GC  &  00  32  47  +39  34 40 &  785   & -10.9  &  35 &  -1.0
\\

Hux C2 &  GC &   00  42  55  +43  57 28 &  785  &  -7.7  &  37 &  ...
\\

NGC 205 & E5pec  & 00  40  22  +41  41 07 &  824   & -16.4  &  40 &
-0.5  \\

And IX &  dSph &  00  52  53  +43  12 00 &  765 &   -8.3 &   42 &  -2.2
\\

And I  &  dSph  &  00  45  40  +38  02 28 &  745  &  -11.8 &   59 &
-1.4  \\

And III  &  dSph  &  00  35  34  +36  29 52 &  749  &  -10.2  &   76 &
-1.7  \\

And V  &  dSph & 01  10  16  +47  37 52  &  774  &  -9.1 &  110 &   -1.9
\\

And X  &  dSph & 01  06  34  +44  48 16  &   783 &   -8.1 &  112 &  -2.0
\\

NGC 147  &  Sph & 00  33  12  +48  30 32  &   675 &  -15.1 &  145 &
-1.1  \\

And II  &  dSph &  01  16  30  +33  25 09  &  652  & -11.8 &  185 &
-1.5  \\

NGC 185 &  Sph & 00  38  58  +48  20 15   &  616  & -15.6 &  190 &  -0.8
\\

M33   &    Sc &  01  33  51  +30  39 37 &   809  &  -18.9 &  208 &  -0.3
\\

And VII  & dSph & 23  26  31  +50  41 31  &   763  &  -12.0 &  219 &
-1.5  \\

IC 10  &   Ir &   00  20  17  +59  18 14  &   825  &  -16.0 &  260 &
-1.3  \\

And VI  &  dSph & 23  51  47  +24  34 57  &   783 &  -11.3 &  269 &
-1.7  \\
 
Pisces & dIr/Sph  & 01  03  53  +21  53 05 &   769 &   -9.8 &  269 &
-1.7  \\

Pegasus &  Ir(?) & 23  28  36  +14  44 35  &   919 &  -12.3 &  474 &
-1.5  \\
 
IC 1613 &  Ir &  01  04  47  +02  08 14  &  700  &  -15.3 &  508 &  -1.3 \\

\enddata
\end{deluxetable}

\begin{deluxetable}{llccrr}
\tablewidth{0pt}   
\tablecaption{ Companions to the Galaxy.}

\tablehead{\colhead{Name}  &\colhead{Type} &  \colhead{RA~~~~~~ Dec}  &
\colhead{D(Gal)}   &  \colhead{$M_{v}$} &\colhead{[Fe/H]}\\
&&\colhead{(J2000)} &\colhead{(kpc)} &&\colhead{(dex)}  } \startdata
N 5139 &  GC   & 13h 26m 46s -47o 28' 37" &   6  &  -10.3 &  -1.6 \\

Sgr   &  dSph  &  18 55 03  -30  28 42 &   19  & -15.0  & -0.5  \\

LMC   &   Ir  &   05 23 35  -69  45  22 &   50 &-18.5  & -0.6  \\

SMC   &   Ir   &  00 52 49  -72  49  43  &  63 &-17.1  &  -1.2  \\

UMi  &   dSph &   15 09 10  +67  12  52  &  69  &-8.9  &  -1.9  \\

Dra  &   dSph  &  17 20 12  +57  54  55  &   79  &-9.4  &  -2.0  \\

Sex  &   dSph &   10 13 03  -01  36  53  &  86  &-9.5  &  -1.9  \\

Scl  &   dSph  &  01 00 09  -33  42  33  &  88  &-9.8  &  -1.5  \\

N 2419 &  GC &    07 38 08  +38  52  55 &   92  &-9.6  & -2.1  \\

Car  &   dSph  &  06 41 37  -50  57  58  &   94  &-9.4  &  -1.8  \\

UMa  &   dSph &   10 34 53  +51  55  12 &  105  &-6.8   & -2  \\

For  &   dSph &   02 39 59  -34  26  57 &  138 &-13.1   &  -1.2  \\

Leo II & dSph  &  11 13 29  +22   9  17 &  205 &-10.1  &  -1.6  \\

Leo I &  dSph  &  10 08 27  +12  18  27  &  270 &-11.9  &  -1.4  \\

Phe  &  dIr/dSph  & 01 51 06  -44  26  41 & 405  &-9.8  &  -1.9  \\

NGC 6822 & Ir   &  19 44 56  -14  52  11 &  500 &-16.0  &  -1.2  \\

\enddata
\end{deluxetable}

\clearpage

\begin{figure}
\caption{Distribution of the distances of the 40 presently known satellites of M31 and the Milky Way system from their parent galaxy.  The six innermost satellites are seen to fall below the relation log R = 1.3 + 0.35 n($<$R), which is plotted in the Figure.  This shows that there appears to be a small excess of satellites at small distances from their parent galaxy.}
\end{figure}

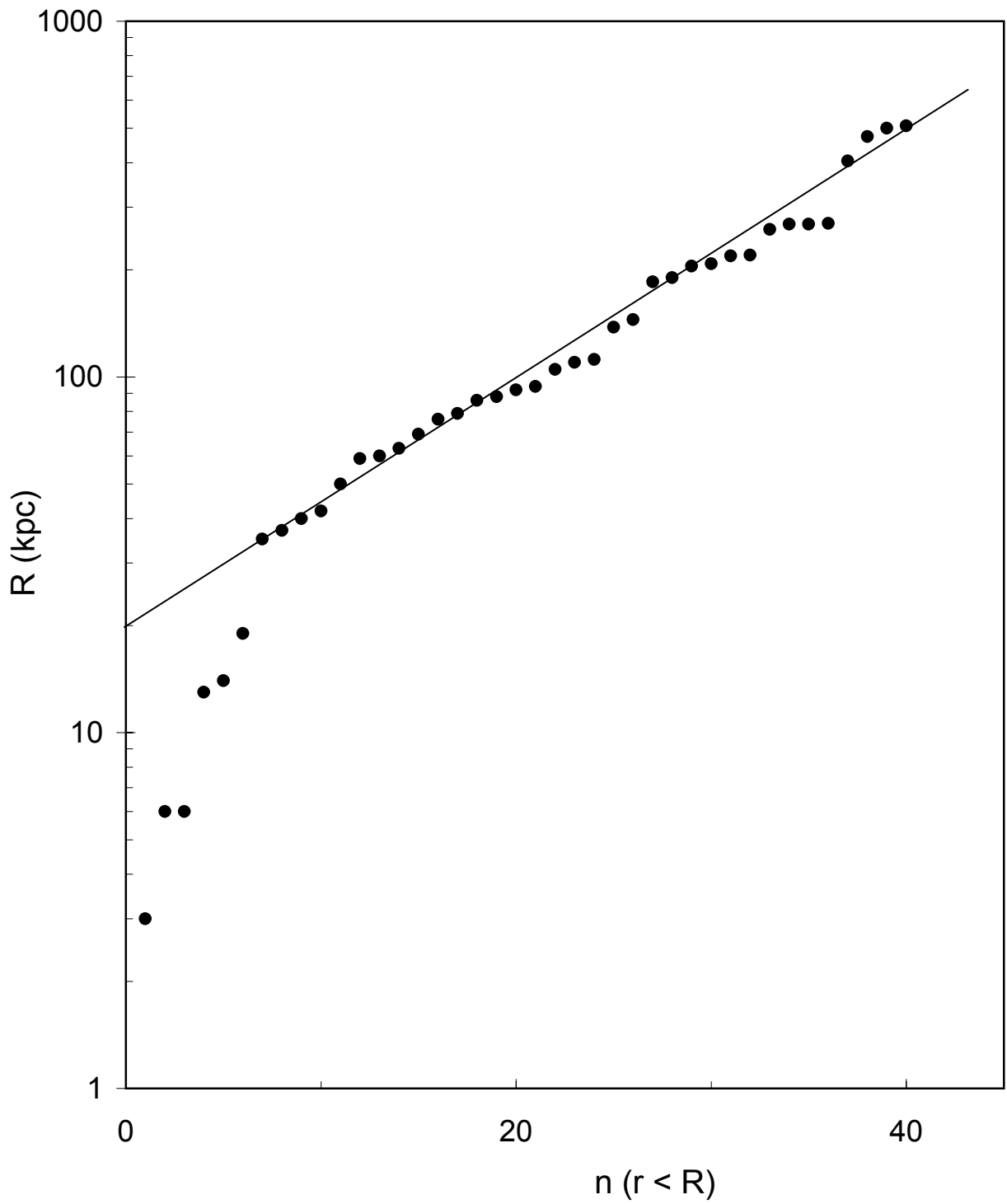
\end{document}